**Ferroelectric nematic liquid crystals: temperature-tunable electrocaloric fluids with large strength at low fields**

**Authors:** Diana I. Nikolova[1*] (corresponding author), Rachel Tuffin[2], Mengfan Guo[3], Neil D. Mathur[3], Xavier Moya[3], Peter Tipping[1], Richard J. Mandle[1,4], Helen F. Gleeson[1]

[1]School of Physics and Astronomy, University of Leeds, Leeds LS2 9JT, United Kingdom
[2]Merck Electronics KGaA, Frankfurter Str. 250, 64293 Darmstadt, Germany
[3]Department of Materials Science, University of Cambridge, Cambridge, United Kingdom
[4]School of Chemistry, University of Leeds, Leeds LS29JT, United Kingdom

*Corresponding author e-mail: phydni@leeds.ac.uk

**Abstract:** The electrocaloric (EC) effect offers a promising energy-efficient and clean cooling technology. We present the first direct measurements of EC temperature change in a new family of EC fluids, ferroelectric nematic liquid crystals (FNLCs), demonstrating in two such materials temperature jumps of $|\Delta T_j|$ ~ 0.2 K for field changes as low as $\Delta E$ ~ 0.1 V μm$^{-1}$. Indirect measurements of adiabatic temperature change $|\Delta T|$ confirm that these direct measurements are an underestimate and that $\Delta E$ = 2 V μm$^{-1}$ can induce up to $|\Delta T|$ ~ 1.6 K, yielding EC strengths $|\Delta T/\Delta E|$ up to 100% higher than incumbent materials. For temperature spans of 5-10 K, we predict a coefficient of performance of ~21-40. We find $|\Delta T|$ ~1 K for >100 FNLCs that collectively span all temperatures between 0˚C and 100˚C. This, together with the new device concepts conceivable with fluid EC materials, offers huge potential for cooling applications.

**Keywords:** electrocaloric effect, liquid crystals, ferroelectric nematic liquid crystals

## 1. Introduction

At the start of the 20th century, the first vapour-compression refrigerators were introduced in America, significantly increasing the average household's energy consumption in comparison to the previously common icebox. At the time, the world's population numbered approximately 2 billion people, a quarter of today's population, and the damaging nature of the refrigerants, both as greenhouse gases (GHGs) and ozone depleting substances, would not be understood for another few decades[1]. A significant amount of work has since been done to address these issues and improve vapour-compression, so it is still the primary technology used in refrigerators, air-conditioning units, and other cooling devices. Nonetheless, cooling accounts for more than 7% of global GHG emissions, and 20% of global energy consumption.[2] Further improvements to this incumbent technology will be neither sufficient in reducing GHG emissions, nor in reducing energy consumption, so research into other refrigeration methodologies is paramount. Fortunately, thermal change can be driven in a few different ways: *via* changes in pressure, uniaxial stress, and application of magnetic or electric fields. These are known, respectively, as the barocaloric, elastocaloric, magnetocaloric, and electrocaloric (EC) effects. It is the latter effect that is of interest in this work as we present a new family of materials which exhibit large EC effects over wide, tuneable temperature ranges and at remarkably low fields. These materials are fluids, rather than solid state materials, and thus additionally offer the potential for automatic energy recovery[3] and new approaches to system design.

In the EC effect, the adiabatic application or removal of an electric field causes a highly reversible temperature change ($\Delta T$) in a material. It can also be viewed as an entropy change ($\Delta S$) under isothermal conditions, and the effect is strongest near phase transitions. For example applying a field in a paraelectric phase reduces the entropy in the system, inducing the more highly ordered ferroelectric phase.[4] Some of the most promising EC materials are ferroelectric ceramics such as $PbSc_{0.5}Ta_{0.5}O_3$ (PST), exhibiting a temperature change of $|\Delta T|$ = 2.2 K for an applied external field $E$ = 2.5 V μm$^{-1}$ measured directly in bulk samples[5–8] and $|\Delta T|$ = 3.6 K in highly-ordered bulk crystals.[9] Notably, the sample geometry matters to the strength of the EC effect that can be realised in a device. This is confirmed by results seen in thin film PST samples[10], and a temperature change as high as $|\Delta T|$ = 5.5 K (at a higher applied field of $E$ = 29 V μm$^{-1}$) in multilayer capacitors (MLCs) using PST.[11] Besides ceramics, a large EC effect has also been observed in copolymers, with $|\Delta T|$ ~ 12.5 K obtained via the indirect method in P(VDF-TrFE) films[12] and as high as $|\Delta T|$ ~ 3.6 K at applied electric fields of $E$ = 100 V μm$^{-1}$ measured using direct infrared measurement in P(VDF-TrFE-CFE) films.[13] Two typical methods for measurements of the EC effect are employed in this work: direct measurements, where $|\Delta T_\mathrm{j}|$ itself is measured; and indirect measurements where $|\Delta T|$ is inferred from polarisation data.

Despite the excellent thermal performance of solid-state ferroelectrics as EC materials, there is still a need for new systems. For example, a longstanding issue in exploiting the EC effect is the presence of extraneous thermal mass in a system – for instance, the necessary addition of heat transfer fluid in a cooling device. Liquid crystals (LCs) are both fluid and highly responsive to applied electric fields, so conceptually, they could function as both the EC material and the heat transfer fluid and be cycled through a system in order to cool it. Some direct and indirect EC data have been published in apolar liquid crystal phases, at transitions between the isotropic (I) phase and a nematic (N) or smectic-A (SmA) phase.[14–17] More recently, indirect measurements in ferroelectric SmC* liquid crystals have been published, indicating a temperature change of $|\Delta T|$ ~ 0.37 K and suggesting ferroelectric liquid crystals are promising EC candidates.[18,19] The possibility of using ferroelectric nematic ($N_f$) liquid crystals as EC materials has also been proposed[20]; these materials have values of polarisation comparable to solid-state materials and are much more fluid than the viscous SmC* systems.[21,22] Here we present the first direct measurements of the EC effect in ferroelectric nematic liquid crystal mixtures, confirming for two such mixtures that the indirect measurements are a good indicator of the exciting EC potential for this new family of materials. We note that extraordinarily small electric fields (0.1 V μm$^{-1}$) are needed, which reduces the chances of dielectric breakdown and is energetically favourable in devices as the lower voltages used can be achieved with simpler electronic circuitry. We finally present indirect data for 129 materials, showing the remarkable tunability of the ferroelectric nematic phase for EC applications.

## 2. Results

Here we present detailed data for two liquid crystal mixtures, FNLC-919 ($N_\mathrm{f} \xleftrightarrow{305.0\ \mathrm{K}} N_\mathrm{x} \xleftrightarrow{317.0\ K} N \xleftrightarrow{353.0\ \mathrm{K}} I$) and FNLC-346 ($N_\mathrm{f} \xleftrightarrow{311.2\ \mathrm{K}} N_\mathrm{x} \xleftrightarrow{317.3\ \mathrm{K}} N \xleftrightarrow{329.0\ \mathrm{K}} I$), provided by Merck Electronics KGaA. Both materials have the phase sequence of ferroelectric nematic ($N_\mathrm{f}$), intermediate ($N_\mathrm{x}$), apolar nematic ($N$), and isotropic ($I$) phases, described in more detail in the Methods section. Zero-field measurements of the entropy change $\Delta S$ at the $N_\mathrm{f}$ to $N_\mathrm{x}$ transition offer an indication of the magnitude of the EC effect that can be achieved in the systems. Both indirect measurements of the EC temperature change $|\Delta T|$ and direct measurements of the EC temperature jumps $|\Delta T_\mathrm{j}|$ are presented; for the two selected liquid crystal mixtures, both sets of measurements are undertaken in electroporation cuvettes with an electrode gap of 1mm, allowing a direct comparison of the two approaches. Indirect measurements on the other 127 mixtures are, however, undertaken in liquid crystal devices with electrode gaps of ~10 μm as the use of conventional liquid crystal devices allows higher fields to be applied where desired and makes the automation of the measurements possible. All the methodology is described in detail in the Methods section.

### 2.1. Indirect Measurements

Indirect measurements of the polarisation and entropy changes of the ferroelectric liquid crystals can be used to give indicative values for the EC temperature change $|\Delta T|$ via the Maxwell method, $|\Delta T| = -\frac{T_\mathrm{S}}{c}\int_{E_1}^{E_2}\left(\frac{\partial P}{\partial T}\right)_E \mathrm{d}E$, as well as EC strength, $|\Delta T/\Delta E|$, and coefficient of performance (COP) of the ferroelectric nematic materials. Calorific measurements of the zero-field transition into the $N_\mathrm{f}$ phase gives us heat flow $dQ/dT$ between the material and its surroundings (shown in Figure 1) and allows us to obtain the associated entropy change, $\Delta S$, which can then be compared to the entropy change under applied field as calculated from the Maxwell method. The transition has a slight hysteresis between heating and cooling for both materials; the average entropy values associated with the transition from the $N_\mathrm{f}$ phase are $\Delta S$ = 1.6 ± 0.1 J K$^{-1}$ kg$^{-1}$ for FNLC-919 and $\Delta S$ = 1.9 ± 0.1 J K$^{-1}$ kg$^{-1}$ for FNLC-346.

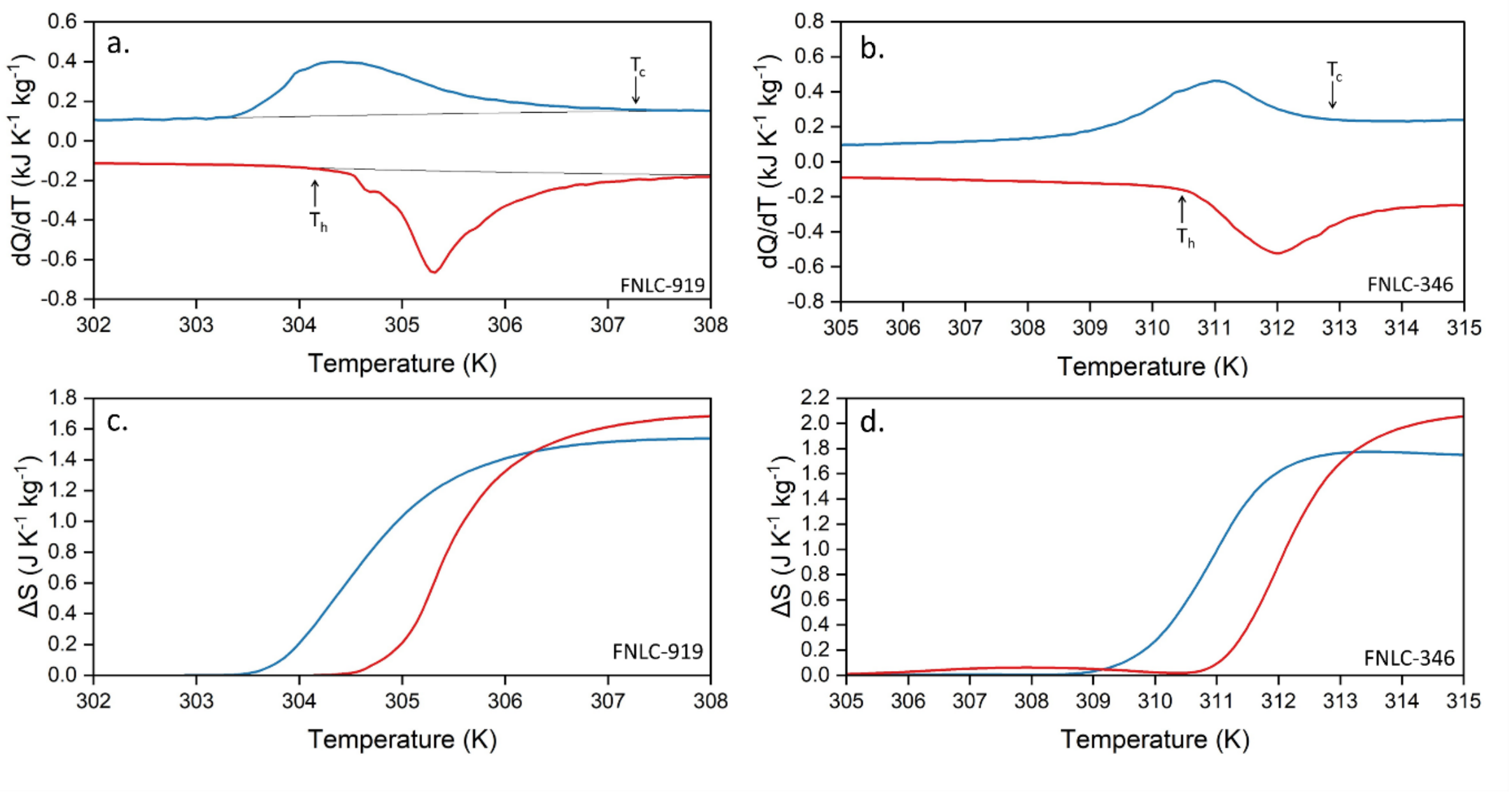

FIG 1. DSC data measured on heating (red) and cooling (blue) at a rate of 10 K/min showing the heat flow dQ/dT across the $N_\mathrm{f} - N_\mathrm{x}$ transition in both materials. (a) Heat flow data for FNLC-919. (b) Heat flow data for FNLC-346. Baselines are shown in black with transition onset points indicated as $T_\mathrm{h}$ and $T_\mathrm{c}$ for heating and cooling respectively. $dQ/dT > 0$ is exothermic. The entropy change determined on heating and cooling across the $N_\mathrm{f} - N_\mathrm{x}$ transition is shown in (c) for FNLC-919 and (d) for FNLC-346.

The polarisation, determined at various field strengths, across the $N_\mathrm{f} - N_\mathrm{x}$ transition, for FNLC-919 and for FNLC-346 is shown in Figures 2 (a) and (c). Using those data, the EC temperature change $|\Delta T|$ can be deduced using the indirect Maxwell method, Figures 2 (b) and (d). In both materials, we see a peak in the temperature and entropy changes close to the $N_\mathrm{f} - N_\mathrm{x}$ transition onset and a shift of this peak towards higher temperatures at larger applied fields. It can be seen that the fields that are required to achieve an appreciable EC effect in these materials are extremely low, around 0.1 V $\mu$m$^{-1}$, which is an order of magnitude lower than is generally necessary in solid state systems.[23] It has recently been shown that the switching in $N_\mathrm{f}$ materials is a function of voltage rather than field[24], meaning that polarisation switching occurs even at very large LC layer thicknesses at the same voltage, and therefore the EC effect can be achieved at extraordinarily low fields. The maximum obtained temperature change in this geometry, suggested by the Maxwell method in FNLC-919 for an applied field $E$ = 0.094 V $\mu$m$^{-1}$, is $|\Delta T|$ = 0.46 K at a temperature of $T_\mathrm{s}$ = 307.8 K. On the other hand, the largest temperature change in FNLC-346 is $|\Delta T|$ = 0.67 K at a temperature of $T_\mathrm{s}$ = 313.8 K at an applied field of $E$ = 0.4 V $\mu$m$^{-1}$.

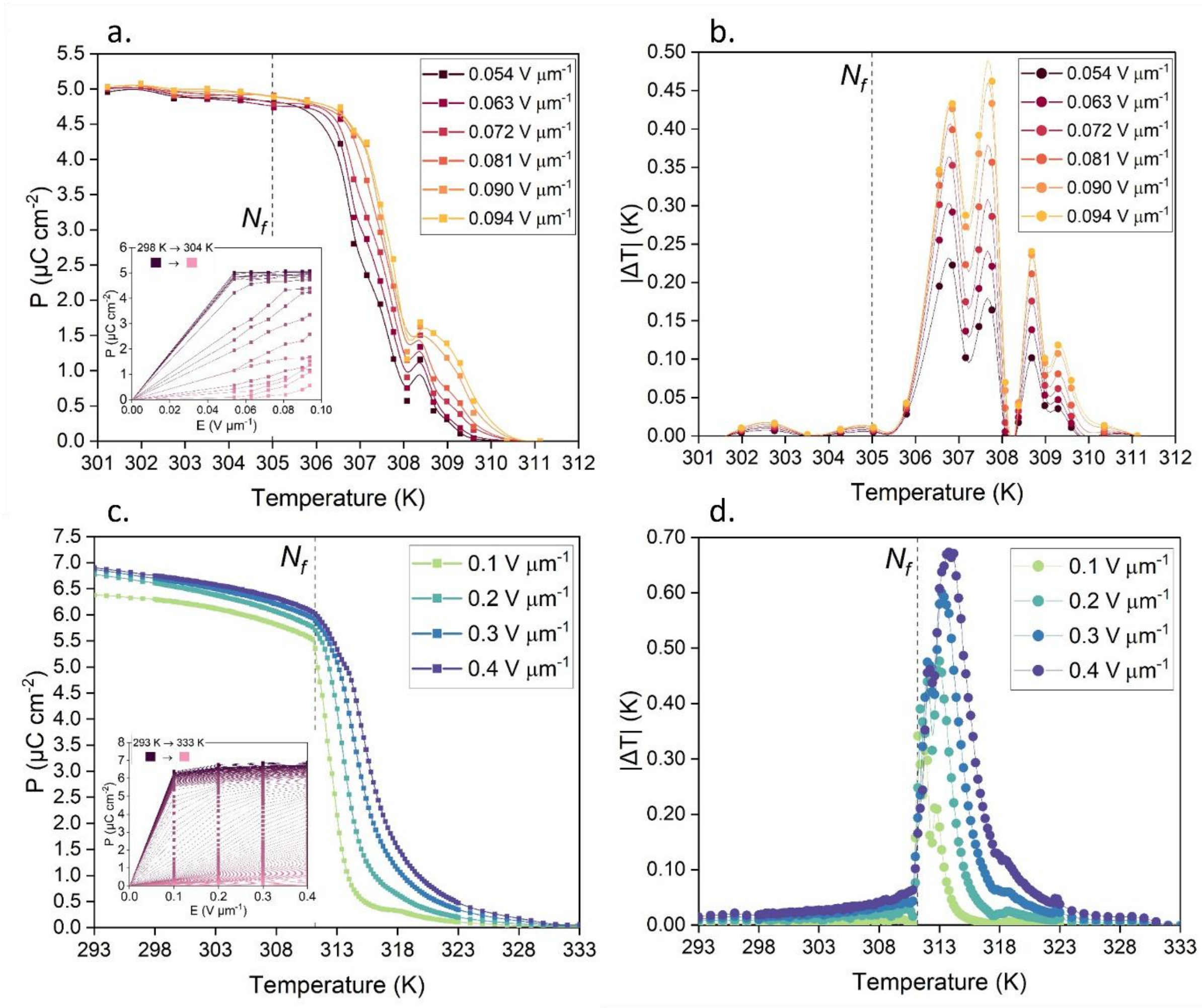

FIG 2. Indirect EC measurements in both materials in cuvettes. Measurements of polarisation $P(T)$ are shown in (a) for FNLC-919 at applied electric fields of 0.054-0.094 V μm$^{-1}$ and (c) for FNLC-346 at applied electric fields of 0.1-0.4 V μm$^{-1}$, with $P(E)$ measurements shown in the smaller inlays. The calculated EC temperature change $|\Delta T|$ is shown for FNLC-919 (b) and FNLC-346 (d). The electrode gap is 1mm in the cuvettes in which the measurements were taken. The ferroelectric transition is indicated on the graphs with a dotted line and labelled.

In both materials, the largest values of $|\Delta T|$ are found slightly above the onset of the $N_\mathrm{f} - N_\mathrm{x}$ transition, a few degrees into the transition range, as expected. The multiple peaks in the $|\Delta T(T)|$ data for FNLC-919 are mostly attributed to the temperature resolution of the polarisation data and the numerical extrapolation that is used to extract the $|\Delta T(T)|$ data. However, the dip in polarisation seen near the $N_\mathrm{f} - N_\mathrm{x}$ transition (at ~308 K) causing the $|\Delta T| = 0$ value is a real and reproducible feature in $P(T, E)$ measurements of ferroelectric nematic liquid crystals, though it is not yet well-understood. Similar observations have been recently published in measurements of the order parameter of these materials[25]. The peak for FNLC-346 is also complex (i.e. a small shoulder can be seen as we move into the $N_\mathrm{x}$ phase), a result of the $N_\mathrm{f} - N_\mathrm{x} - N$ transitions being much closer together than in FNLC-919. Appreciable temperature changes can also be seen just below the onset of the $N_f$ transition for FNLC-346; at the lowest field strength of $E$ = 0.1 V μm$^{-1}$ we find $|\Delta T|$ = 0.34 K at $T_\mathrm{s}$ = 311.2 K.

From the polarisation measurements we also calculate entropy change values of $\Delta S$ = 2.05 J K$^{-1}$ kg$^{-1}$ in FNLC-919 and $\Delta S$ = 2.79 J K$^{-1}$ kg$^{-1}$ in FNLC-346 at the respective maximum applied fields (see Figure S1 in the SI for the full entropy change graphs). These are both larger than the zero-field entropy change values measured using DSC, shown previously in Fig. 1. This is particularly interesting as the strongest EC effect is not achieved across the entire $N_f - N_x$ transition, which spans a range of ~4 K in both materials, but instead 1-2 K into the transition, closer to its peak. Over the temperature range of the transition, the $N_f$ and $N_x$ phases may coexist in a small biphasic region as is common in mixtures. This means that the application of electric field only induces a transition in the $N_x$ proportion of the mixture, not its entirety. Despite this, a large entropy change is achieved due to the combined contributions of the phase transition and the electrocaloric response – even larger than the zero-field entropy change spanning the entire transition which was shown in Fig. 1.

In all the indirect measurements presented above, made using the cuvettes, the applied fields were < 0.4 V μm$^{-1}$. It is, however, routine to apply significantly higher fields in conventional liquid crystal devices which have electrode gaps of ~10 μm. For the materials considered here, at $E$ = 2 V μm$^{-1}$ (achieved in a conventional device), the predicted temperature change can be as high as $|\Delta T|$ = 1.2 K (see Figures S2 and S3 in SI II for data). The full potential of the EC temperature change in other ferroelectric nematic materials can be seen by considering indirect measurements made in such devices.

### 2.2. Direct Measurements

Ideally, a direct measurement of the temperature change $|\Delta T_j|$ upon application of an electric field at a set temperature $T_s$ is desired to ascertain the magnitude of the EC effect of a material. Importantly, the fact that such measurements show both heating *and* cooling confirm the EC effect; measurements of heating alone could be attributed to other phenomena such as Joule heating.

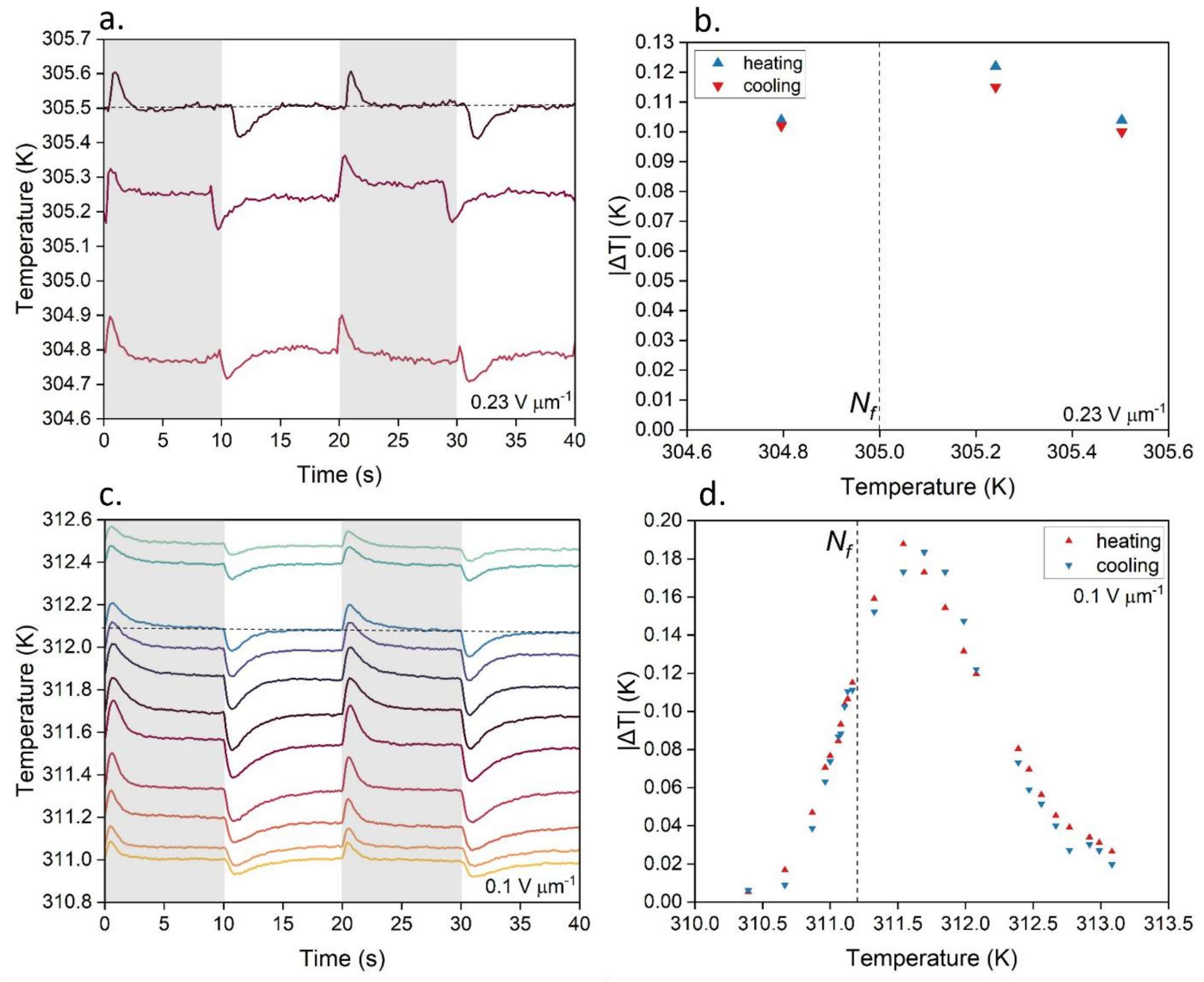

FIG 3. Direct EC measurements in both materials. (a) The direct EC effect in FNLC-919 at $E$ = 0.23 V μm$^{-1}$ with (b) achieved temperature changes (cooling and heating) shown in the $\Delta T_{\mathrm{j}}(T_{\mathrm{s}})$ plot adjacent. (c) Direct temperature measurements in mixture FNLC-346 and (d) the associated temperature change $\Delta T_{\mathrm{j}}(T_{\mathrm{s}})$ at $E$ = 0.1 V μm$^{-1}$. All measurements were taken in cuvettes with electrode gap of 1mm, the field-on and -off cycle is indicated by shaded and clear areas respectively in (a) and (c) and selected horizontal lines show the baseline temperature.

The EC heating and cooling measured for the ferroelectric nematic materials is shown in Figure 3 and summarised in Table 1. Tests in FNLC-919 were performed at $E$ = 0.23 V μm$^{-1}$, shown in Figure 3 (a-b). The largest heating in FNLC-919 is $+\Delta T_{\mathrm{j}}$ = 0.12 K achieved at $T_{\mathrm{s}}$ = 305.2 K, slightly above the onset of the transition. Maximum cooling is achieved at the same temperature, also with magnitude $-\Delta T_{\mathrm{j}}$ = 0.12 K. Figure 3(c-d) shows the directly measured heating and cooling from the EC effect in FNLC-346 with an applied electric field $E$ = 0.1 V μm$^{-1}$. The maximum effect occurs at set temperature $T_{\mathrm{s}}$ = 311.5 K with a temperature increase (heating) $+\Delta T_{\mathrm{j}}$= 0.19 K and a corresponding decrease (cooling) of $-\Delta T_{\mathrm{j}}$= 0.17 K.

These $|\Delta T_{\mathrm{j}}|$ values are 2-3 times smaller than the $|\Delta T|$ values predicted via indirect measurements for a similar applied electric field due to the presence of inactive thermal mass in the system. A closer correspondence was achieved in FNLC-346 than in FNLC-919, where the multiple phase transitions that are farther apart in temperature affect the system. Multiple peaks in the directly measured $|\Delta T_{\mathrm{j}}(T)|$ data are not observed as they were in the indirect $|\Delta T(T)|$ results as most were artefacts of the numerical methods employed and the resolution of the apparatus used for direct measurements was too low to observe the 'dip' predicted at 308 K.

The heating and cooling achieved on each cycle is remarkably symmetric, and there is negligible Joule heating in these materials (≤ 0.01 K). As shown elsewhere[26], Joule heating in direct EC measurements manifests as a change in the baseline temperature, which was not observed here, as indicated by the flat dotted lines in Fig. 3(a) and (c). The response time for heating is typically around 0.5 s, while for cooling it is 0.7 s. In particular, there is a delay associated with the thermocouple response time, which is slower than the response time of the liquid crystal. That, combined with the large amount of inactive thermal mass present in the system (i.e. the electrodes, thermocouple, and unaddressed fluid), means that the direct values are expected to be an underestimate of the actual temperature change.

The temperature range of the EC effect is an important feature of any material that is being considered for applications. For instance, although measurements of the non-polar liquid crystals 12CB[15] and 14CB[16] suggest a significant EC temperature change might be expected near the first order smectic-A to isotropic transition, the effect occurs over a range of less than 0.5 K around the set temperatures, which is not useful for practical applications. In the ferroelectric nematic materials reported here, the temperature range over which at least 50% of the maximum $|\Delta T|$ can be achieved gives a further indication of the practical applicability of these materials in EC systems. For both of our materials, this range is ~2.5 K. By further optimising the system (for example, by altering the sample geometry as in the case of using MLCs[27] rather than measuring bulk ceramic samples) and exploiting the active energy regeneration capabilities of the liquid crystals, this range could be increased.

| Material | $T_{\mathrm{s}}$ (K) | Field for indirect measurement $\lvert E\rvert$ (V μm$^{-1}$) | Indirect $\lvert\Delta T\rvert$ (K) | Field for direct measurement $\lvert E\rvert$ (V μm$^{-1}$) | Direct $\pm\Delta T_{\mathrm{j}}$ on cooling/heating (K) | EC strength $\lvert\Delta T\rvert/\lvert\Delta E\rvert$ (K μm V$^{-1}$) |
|---|---|---|---|---|---|---|
| FNLC-919 | 305.2 | 0.094 | 0.46 | 0.23 | 0.12 / 0.12 | 4.89 |
| FNLC-346 | 311.2 | 0.1<br>0.4 | 0.34<br>0.67 | 0.1 | 0.17 / 0.19 | 3.40<br>1.67 |

**Table 1.** Summary of the direct and indirect measurements of the EC effect in the cuvette geometry. Two values are shown for FNLC-346, one at the highest achieved temperature change $|\Delta T|$ and one at the highest achieved EC strength $|\Delta T/\Delta E|$.

## 2.3. **Electrocaloric strength and COP**

Neither the entropy change nor the temperature change associated with the EC effect can exist in a proverbial vacuum; the applied electric field strength needed to achieve a specific temperature change is also important; lower fields are desired both to save energy and because the associated electronics in both measurements and systems are far simpler to devise. Therefore, we also calculate the EC strength $|\Delta T|/|\Delta E|$ of the ferroelectric nematic materials, which allows both a direct comparison with other EC materials and a figure of merit that suggests their potential for real world applications.

The largest EC strength $|\Delta T|/|\Delta E|$ exhibited by these materials, deduced from the indirect method (in cuvettes) is 4.89 K μm $V^{-1}$ for FNLC-919 and 3.40 K μm $V^{-1}$ for FNLC-346. Given the inactive thermal mass which suppresses direct measurements, it was determined the indirect measurements are more reliable for calculating EC strength and more useful for comparison. Such a comparison with indirect measurements of materials from the literature is shown in Figure 4a. The material FNLC-919 in particular shows a remarkable improvement over existing materials.

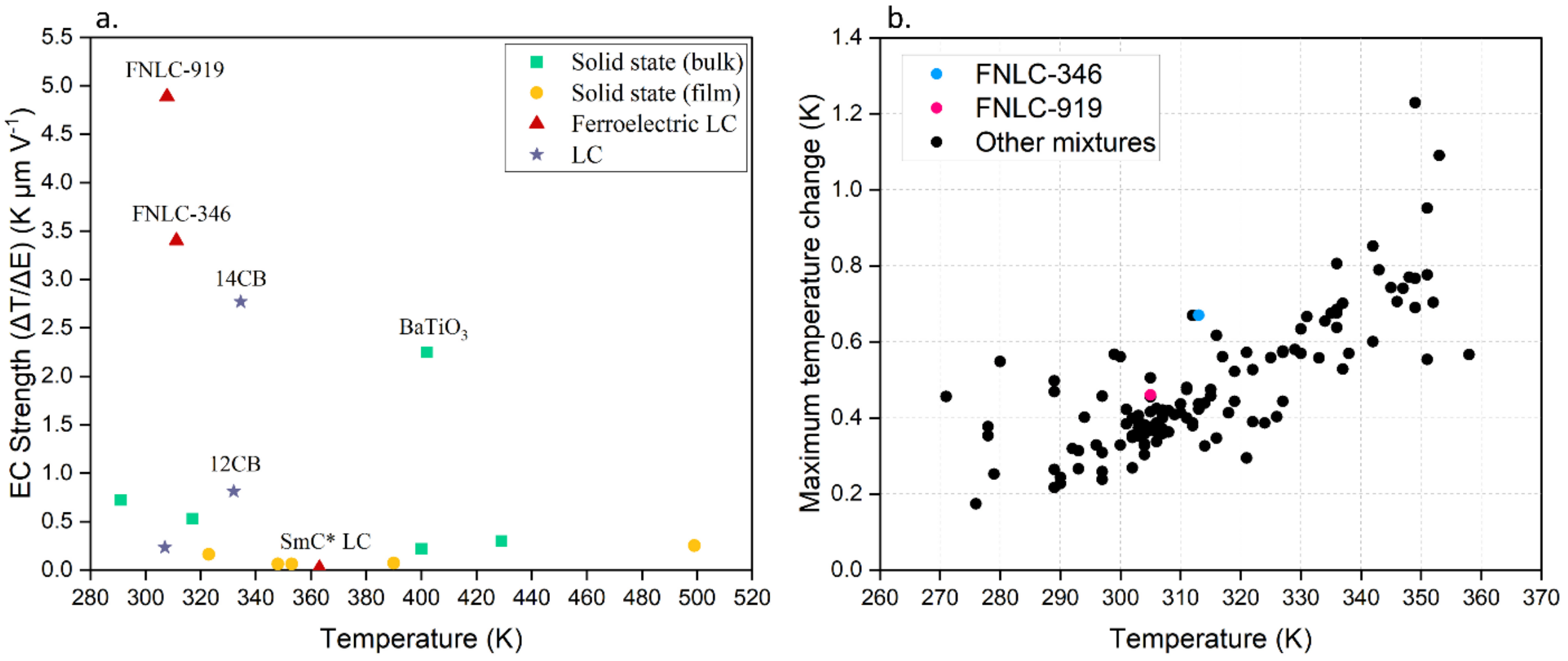

FIG 4. EC strength comparison graph and ECE tunability graph. (a) Comparison of indirect EC strength values for the materials examined in this paper (at $E \leq 0.4$ V $\mu m^{-1}$ as previously described) to published values from the literature.[10–13,15,17,18,28] The materials 12CB and 14CB[16], $BaTiO_3$[29] and a chiral smectic liquid crystal material[19] are highlighted. (b) $|\Delta T|$ values for 129 liquid crystal mixtures which exhibit the ferroelectric transition at a range of temperatures from 271 K to 358 K, made using the indirect method at $E = 1$ V $\mu m^{-1}$. The indirect temperature change data from this work for FNLC-919 and FNLC-346 are included on this graph as red and blue points respectively for comparison.

It has already been mentioned that the temperature range of the EC effect in ferroelectric nematic liquid crystals is promising. A further advantage is that the phase transitions of liquid crystals are highly tunable, allowing the ferroelectric to paraelectric transition to be engineered in different mixtures to take place at any desired temperature over a wide temperature range. Indeed, Figure 4b shows the phase transition temperatures and $|\Delta T|$ values measured independently via the Maxwell method for 129 ferroelectric nematic mixtures designed by Merck Electronics KGaA (Darmstadt, Germany) [20] in which the ferroelectric to paraelectric transition may be anywhere from 271 K to 358 K, showcasing an EC temperature change of 0.2-1.2 K for an applied field of $E = 1$ V $\mu m^{-1}$. This allows huge flexibility in terms of potential applications.

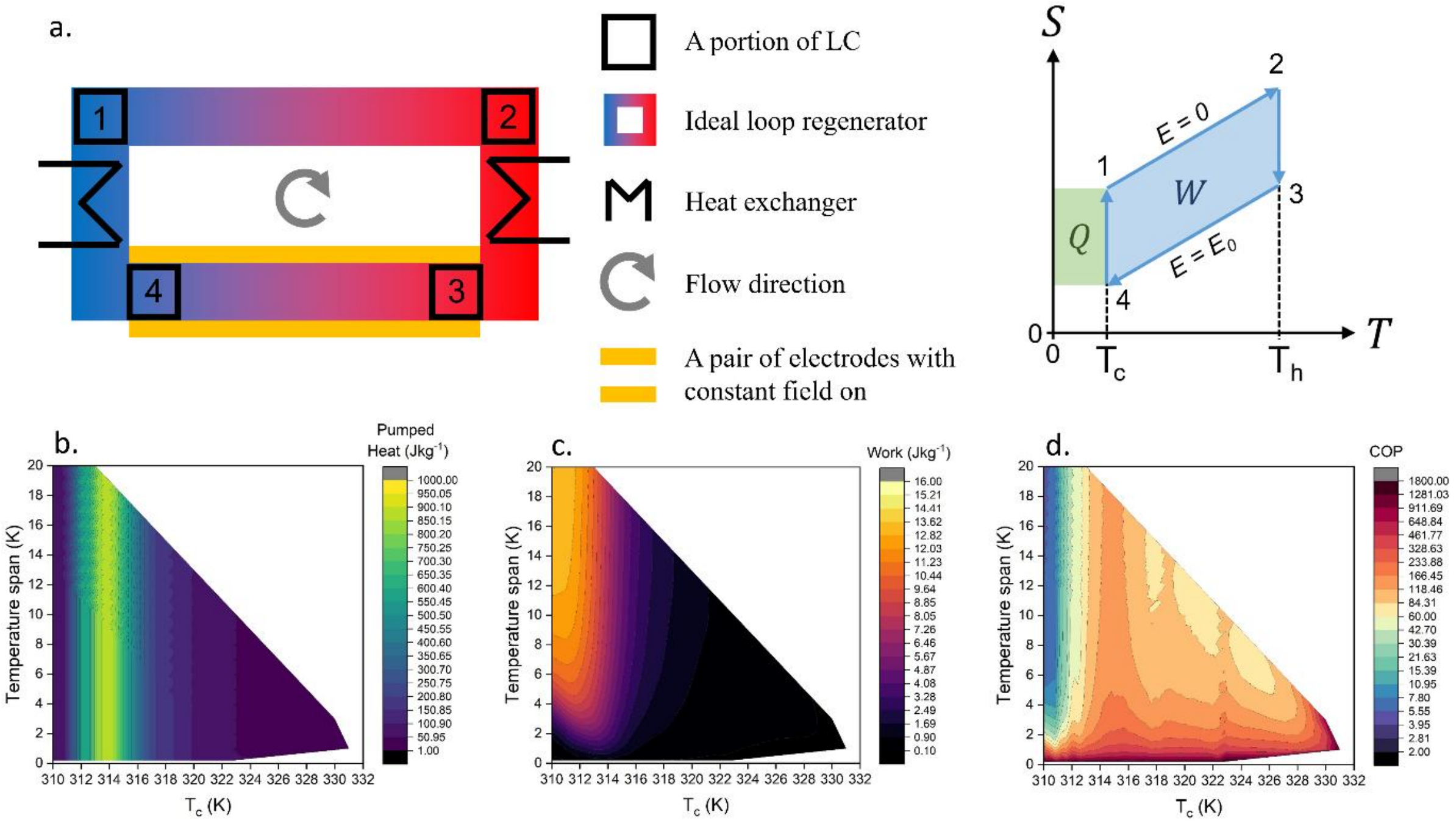

FIG 5. Coefficient of performance schematic and data for FNLC-346. (a) Schematic of a portion of liquid crystal moving through an ideal regenerator loop, and the corresponding Ericsson cycle showing the pumped heat $Q$ and work $W$ done at the cold end temperature $T_c$. (b) Colour map of pumped heat $Q$ values for FNLC-346 at the maximum applied field of 0.4 V μm$^{-1}$. (c) Colour map of work $W$ done for FNLC-346. (d) Colour map of COP values for FNLC-346 obtained via $COP = Q/W$.

The indirect measurements of entropy change (shown in SI I) were used to obtain coefficient of performance (COP) values assuming an ideal regenerator to illustrate the potential of these materials for EC cooling applications. The regenerator is with a loop shape where the liquid crystal moves in a clockwise direction without heat leakage, and an Ericsson cycle (see Figure 5a) can be driven for a part of the liquid crystal during its run through the whole loop. The area contained within the four legs of the Ericsson cycle represents the work done during the process, $W$, where we assume electrical energy recovery (so $W$ corresponds to the cycle area rather than the gross electrical charging work); the pumped heat at the cold end is an integral taken with respect to the entropy, as shown in Figure 5a. We obtain the COP via $\text{COP} = Q/W$. As finite power-electronics losses, dielectric losses, leakage, heat-transfer losses, and other device-level parasitics are not included in the calculation, this provides an upper-bound COP value for the Ericsson cycle. Colour maps of the pumped heat and work values for FNLC-346 are shown in Figure 5b and 5c respectively. Colour maps of the resulting COP values are shown in Figure 5, with respect to the temperature spans corresponding to each value. Such colour maps are shown for FNLC-919 in Figure S5 in SI IV. Taking a useful temperature span of 5-10 K, we see COP ~ 21-40 near the ferroelectric transition. These values are comparable to reported COPs in the literature - for instance, Crossley et al.[8] report COP = 12-27 for a temperature span of 10 K in PST and clamped PST. Interestingly, at higher temperatures, i.e. 315 K, for the temperature span of 10 K we observe a COP as high as 80.

## 3. Discussion

In this work, we have shown the first direct measurements of the EC effect in ferroelectric nematic liquid crystals, demonstrating their potential in EC devices and applications. Two materials were considered in detail, evaluated first via the indirect Maxwell method and then via direct measurement. Both exhibit extremely symmetric heating and cooling cycles and very large EC strengths $|\Delta T/\Delta E|$ near room temperature; indeed, one of the materials discussed (FNLC-919) has a strength that exceeds the maximum value reported for other systems by at least 100%. The electric fields at which the measurements were made are an order of magnitude lower than in solid-state systems, a factor that can appreciably simplify the electronics that would be needed to drive an EC heat pump. Based on the EC entropy change, upper-bound COP values as high as 21-40 for temperature spans of 5-10 K are predicted in ferroelectric nematic liquid crystals. Additionally, the EC effect is achieved over a temperature range where at least 50% of the maximum $|\Delta T|$ is maintained over a temperature range of ~ 2.5 K, larger than any other liquid crystal system. The indirect measurements predicted a larger temperature change $|\Delta T|$ than was directly measured in both materials, though it was expected that the direct temperature change $|\Delta T_j|$ values would be an underestimate due to the large amount of inactive thermal mass present. By minimizing the latter, the technique can be improved and better agreement with the indirect prediction can be achieved. This is the subject of further work.

Further, we have demonstrated through indirect measurements that ferroelectric nematic liquid crystals, in which the ferroelectric phase transition can be engineered to take place at any desired temperature between 0˚C and 100˚C, have unprecedented tunability of the EC effect. In addition to their comparable or improved EC strength with respect to incumbent materials, the liquid nature of the ferroelectric nematic materials would allow the possibility of them being used as the regenerative fluid in a system, allowing one to conceptualise completely new approaches to EC system design[3] and opening up a wide variety of possible applications.

## 4. Methods

### 4.1. Materials

The materials examined here in detail are FNLC-346 ($N_f \xleftrightarrow{311.2\text{ K}} N_x \xleftrightarrow{317.3\text{ K}} N \xleftrightarrow{329.0\text{ K}} I$) and FNLC-919 ($N_f \xleftrightarrow{305.0\text{ K}} N_x \xleftrightarrow{317.0\,K} N \xleftrightarrow{353.0\text{ K}} I$), thermotropic liquid crystal mixtures which exhibit the ferroelectric nematic ($N_f$) phase at room temperature. In addition, these mixtures have two higher-temperature nematic phases, the usual non-polar nematic ($N$) phase and an intermediate nematic phase reported to have antiferroelectric character, here denoted as $N_x$, but elsewhere also called a $N_s$[30] or $SmZ_A$[31] phase. All materials are provided by Merck Electronics KGaA, Darmstadt, Germany and the values for the phase transitions were obtained via differential scanning calorimetry (DSC).

### 4.2. Direct and indirect measurement of the electrocaloric effect

There are a number of difficulties when it comes to direct measurement of a fluid material – chiefly the fact that the fluid must be in confinement to allow consistent field application. Usually in the case of LC systems, the material is confined within a device in which the substrates are glass, which has a low thermal conductivity, the electrodes are indium tin oxide (ITO) and the dimensions are such that the glass is ~1mm thick while the LC layer is ~10 μm thick. Such a confined geometry with such a small proportion of the EC material (the liquid crystal in an average glass cell takes up about ~0.5% of the total cell volume, the other 99.5% being glass) is problematic for direct measurements of temperature changes and as a result, one of the most common ways to evaluate liquid crystalline materials for their EC effect is through indirect measurements. We first outline that approach, then describe how direct measurements were achieved for the ferroelectric nematic materials.

The most straightforward way to estimate the potential EC effect in a material is through zero-field differential scanning calorimetry (DSC) measurements. In this work, DSC was used to obtain heat flow $(dQ/dT)$ curves across the ferroelectric transition in each material, both on heating and cooling. From this, zero-field entropy change $\Delta S$ across the ferroelectric transition was found as the integral of $(dQ\ /\ dT)/T$. This zero-field entropy estimate can then be compared to the entropy change deduced from the Maxwell method, as shown below.

The indirect Maxwell method, stemming from the Maxwell relation $\left(\frac{\partial S}{\partial E}\right)_T = \left(\frac{\partial P}{\partial T}\right)_E$, also offers a convenient approach to estimate $\Delta S$ and $|\Delta T|$ in ferroelectric materials through equations (1) and (2),

$$\Delta S = \int_{E_1}^{E_2} \left(\frac{\partial P}{\partial T}\right)_E \mathrm{d}E\,, \qquad (1)$$

and

$$|\Delta T| = -\frac{T_s}{c}\int_{E_1}^{E_2}\left(\frac{\partial P}{\partial T}\right)_E \mathrm{d}E. \tag{2}$$

Here, $P$ is the polarization of the ferroelectric material, $E_1$ and $E_2$ are the initial and final applied electric fields respectively, $T_s$ is the set temperature, $c$ is the volumetric heat capacity. $\Delta S$ is often reported in addition to $|\Delta T|$. The use of an entropy map methodology as presented by Mathur *et al.*[8,11] takes into account the fact that the heat capacity can vary with temperature change and on application of a field.

To make indirect measurements of $|\Delta T|$, the polarization $P$ is measured isothermally at selected set temperatures $T_s$ for a number of applied electric fields, $E$, at each temperature. This indirect measurement approach in ferroelectric LCs is described in detail in ref [19]; conventional LC cells are employed with the thickness of the LC layer typically being ~5-50 μm. It is noteworthy that these typical LC layer thicknesses mean that the fields applied are usually ~1-5 $\mathrm{V\mu m^{-1}}$, an order of magnitude higher than is applied in the cuvettes used for the direct measurements described below. The polarisation measurement is undertaken using the current reversal method with a triangular AC waveform at a certain frequency (in this case, 20 Hz), as described by Miyasato et al[32] and the polarisation is determined through $P = \int \frac{i}{2A}\mathrm{d}t$ (where $i$ is the polarisation current measured as the field is reversed and $A$ is the electrode area). In this work, both conventional cells (more detailed discussion in Supplementary Note I) and the cuvettes used for the EC measurements (described below) were used for indirect measurements, the latter allowing a direct comparison between the direct and indirect measurements.

There is no published method for obtaining direct EC measurements in a fluid ferroelectric material, and the fluid nature of the liquid crystals means that methodologies commonly used for evaluation of solid-state materials (e.g. contact thermometry) are difficult to employ. As such, a bespoke experimental set-up was devised in which a small amount (< 1 ml of addressed material) of the liquid crystal is contained in an electroporation cuvette, chosen for its built-in aluminium electrodes with 1-4 mm gaps (further details of the cuvettes are given in Supplementary Note II). An electrically insulated thermocouple is submerged directly into the liquid crystal fluid, between the electrodes (see Figure 6a).

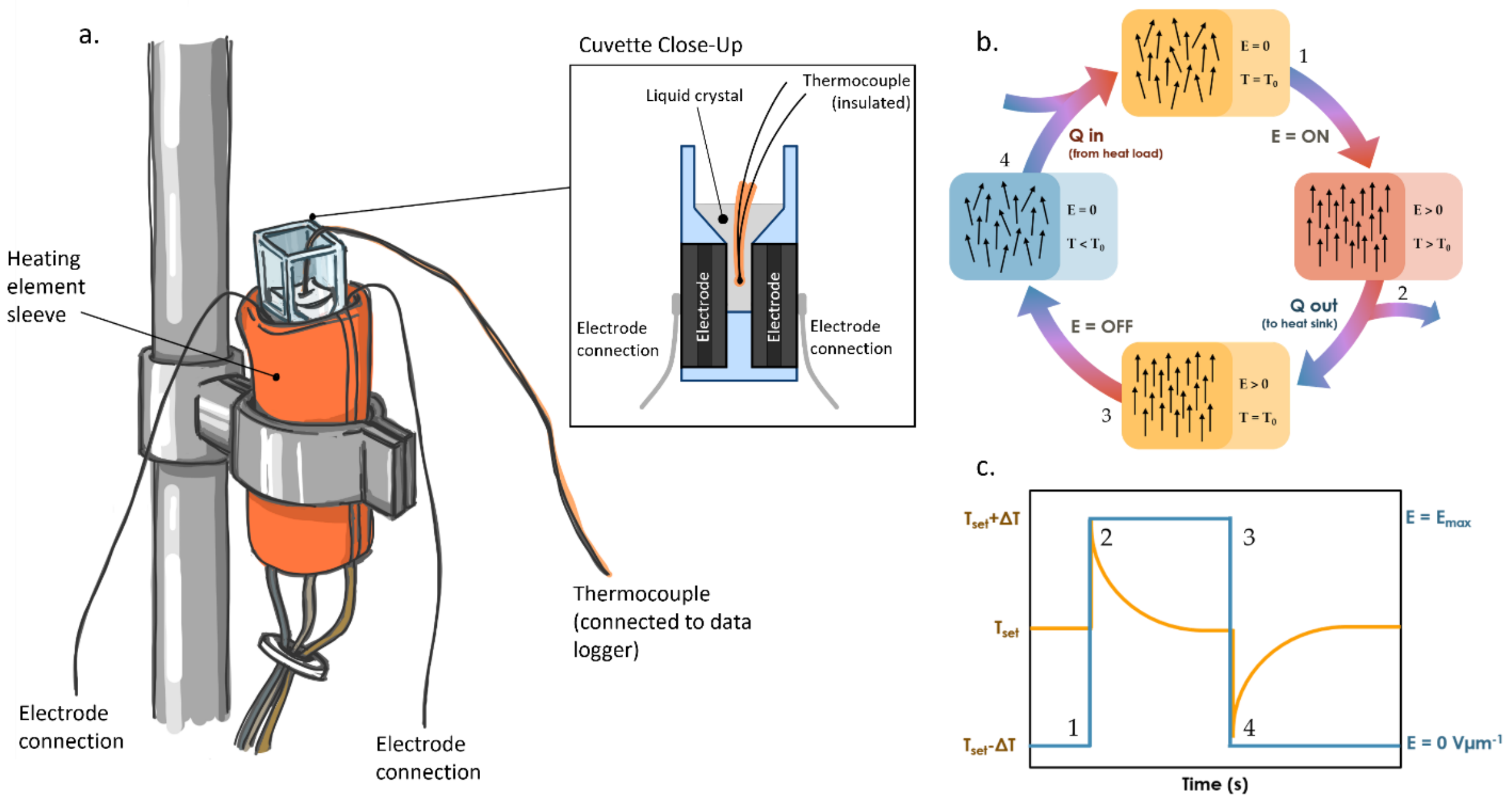

FIG 6. Direct EC measurement set-up schematics and ECE diagrams. (a) Illustration showing the experimental set-up for both direct and indirect EC measurements, showing the electroporation cuvette held in a heating sleeve. The inlaid close-up diagram shows how the cuvette electrodes are connected as well as the insertion of the electrically insulated thermocouple (insulating layer depicted in orange), submerged in the liquid crystal. (b) A diagram of the four-step EC cycle, showing the interaction between changing the entropy associated with orientational order by applying or removing electric field and the subsequent change in thermal vibrations. (c) An idealised schematic of the direct measurement; a field is applied at 1 and removed at 3; the corresponding heating (2) and cooling (4) peaks are shown in orange.

The sample is held in a temperature-controlled oven for the duration of the experiment allowing the magnitude of the EC effect to be determined over a range of temperatures. For direct EC measurements, a DC field is applied, then removed, while temperature data from the thermocouple (technical details for the thermocouples can be found in SI II) are collected via a data logger or oscilloscope. The applied voltage was limited in this experimental arrangement, constraining the maximum the applied field to 0.1 V $\mu m^{-1}$. The direct measurements taken in cuvettes are shown in Figure 6(b-c), following the four-step EC cycle, with each measurement being undertaken at a set electric field strength $E$ and temperature $T_s$. The steps of the cycle are as follows: (1) An adiabatic jump in temperature $|\Delta T_j|$ upon application of the field ($E_{on}$), (2) a slow, isofield reduction in temperature back to the set temperature $T_s$, (3) an adiabatic drop in temperature $\Delta T_d$ upon removal of the field ($E_{off}$), and (4) a slow return to the set temperature $T_s$ with no field applied.

**Acknowledgements.** We are grateful for funding from the EPSRC, grant no. EP/V054724/1. R.J.M. Acknowledges funding from UKRI via a Future Leaders Fellowship, grant no. MR/W006391/1. We also acknowledge Merck Electronics KGaA for funding and the provision of materials. M.G. was supported by a Newton International Fellowship from the Royal Society and a Goldsmiths' Early Career Research Fellowship from Churchill College, Cambridge.

**Author contributions.** DSC measurements and related calculations were performed by D.I.N. Polarisation measurements and direct electrocaloric measurements were performed by D.I.N. and R.T. Direct method was devised by D.I.N. with input from all authors. COP calculations were performed by M.G. Manuscript was written by D.I.N., R.T., M.G., and H.F.G., with guidance and input from P.T., R.J.M., X.M., and N.D.M.

**Competing interests.** The authors declare no competing financial or non-financial interests.

**Data Availability.** The data associated with this article can be found at DOI: https://doi.org/10.5518/1739